\documentclass[12pt]{article}

\usepackage{epsfig}
\usepackage{amssymb}
\usepackage{amsmath}
\usepackage{amsfonts}
\usepackage{graphicx}
\usepackage{mathrsfs}
\usepackage[dvips]{color}
\usepackage{multirow}
\usepackage{lipsum}

\newcommand{\fz}{\mathfrak{z}}

\newcommand{\fK}{\mathfrak{K}}

\newcommand{\bH}{\mathbf{H}}
\newcommand{\bI}{\mathbf{I}}

\newcommand{\bM}{\mathbf{M}}

\newcommand{\bU}{\mathbf{U}}

\newcommand{\cK}{\mathcal{K}}

\newcommand{\be}{\begin{equation}}
\newcommand{\ee}{\end{equation}}
\newcommand{\bea}{\begin{eqnarray}}
\newcommand{\eea}{\end{eqnarray}}
\newcommand{\nn}{\nonumber}

\newcommand{\ed}{\end{document}}

\newcommand{\np}{\newpage}

\newcommand{\bi}{\begin{itemize}}
\newcommand{\ei}{\end{itemize}}

\newcommand{\bce}{\begin{center}}
\newcommand{\ece}{\end{center}}

\newcommand{\RE}{{\rm Re}}
\newcommand{\IM}{{\rm Im}}

\newcommand{\bzero}{{\mathbf{0}}}

\begin{document}

\title{Perfect Broadband Invisibility in Isotropic Media\\ with Gain and Loss}

\author{Farhang Loran\thanks{Email Address: loran@cc.iut.ac.ir}\:  and  Ali~Mostafazadeh\thanks{Corresponding author, Email Address: amostafazadeh@ku.edu.tr}\\[6pt]
$^*$Department of Physics, Isfahan University of Technology,\\ Isfahan 84156-83111, Iran\\[6pt]
$^\dagger$Departments of Mathematics and Physics, Ko\c{c} University,\\ 34450 Sar{\i}yer,
Istanbul, Turkey}

\date{ }
\maketitle

\begin{abstract}

We outline a method for constructing effectively two-dimensional isotropic optical media that are perfectly and omnidirectionally invisible for both TE and TM waves provided that their wavenumber does not exceed a preassigned value $\alpha$. This relies on the observation that a complex scattering potential $v(x,y)$ displays perfect invisibility for wavenumbers $k\leq\alpha$ provided that its Fourier transform with respect to $y$, which we denote by $\tilde v(x,\fK_y)$, vanishes for $\fK_y\leq 2\alpha$. We use this result to construct the permittivity profile for a realistic optical slab that is perfectly invisible in the wavenumber window $[0,\alpha]$.
\vspace{2mm}



\end{abstract}

Can a scattering potential be invisible in a spectral band of arbitrarily large width? This is a natural question of great theoretical and practical importance. The use of conformal mappings \cite{leonhardt}, metamaterials \cite{pendry-2006,schuring-2008}, and especially crafted anisotropic material \cite{rahm} has led to some remarkable progress in this subject. But achieving perfect (non-approximate) broadband invisibility for ordinary nonmagnetic isotropic material has remained out of reach. This letter proposes a simple method for accomplishing this goal.

In one dimension, if a real or complex potential $v(x)$ decays exponentially (or more rapidly) as $x\to\pm\infty$, the reflection amplitudes are complex-analytic functions of the wavenumber $k$, \cite{blashchak}. Because a nonzero complex-analytic function can vanish only at a set of isolated points in complex plane, $v(x)$ can be reflectionless either in the entire frequency spectrum (full-band) or at a discrete set of isolated values of the frequency. This means that reflectionlessness and invisibility in a spectral band of finite width (finite-band) are forbidden for such short-range potentials.

The problem of finding full-band reflectionless real potentials in one dimension has been addressed in the 1950's \cite{kay-moses}. The outcome is a class of potentials with an asymptotic exponential decay, which have recently found applications in designing antireflection coatings \cite{gupta-2007,thekkekara-2014}.

For a complex scattering potential, the reflection coefficients for the left and right incident waves need not be equal. In particular it can be invisible from one direction and visible from the other \cite{lin}. This observation has recently attracted a lot of attention and led to a detailed study of the phenomenon of unidirectional invisibility \cite{invisible3-1,pra-2013a,invisible3-2,invisible3-3,pra-2014a,pra-2015,horsley,longhi-epl, longhi-ol-2015,longhi-ol-2016,mustafa}.

Consider the Schr\"odinger equation $-\psi''+v(x)\psi=k^2\psi$ for a complex potential of the form:
    \be
    v(x)=\chi_a(x)f(x),~~~~
    \chi_a(x):=\left\{\begin{array}{cc}
    1 &{\rm for}~x\in[0,a],\\
    0 &{\rm otherwise},\end{array}\right.
    \label{e1}
    \ee
where $f(x)$ is a periodic potential with period $K:=2\pi/a$. We can express $f(x)$ in terms of its Fourier series, $f(x)=\sum_{n=-\infty}^\infty c_n e^{i nKx}$. It turns out that if $v(x)$ is sufficiently weak, so that the first Born approximation is reliable, and  $c_0=c_{-m}=0\neq c_m$ for some integer $m>0$, then $v(x)$ is unidirectionally invisible from the left for the wavenumber $k=\pi m/a$, \cite{pra-2014a}. The simplest example is $f(x)=c_1 e^{iKx}$ whose investigation led to the discovery of unidirectional invisibility \cite{invisible-1,invisible-2,invisible-3,lin}.

As noted in Ref.~\cite{pra-2014a}, if $c_n=0$ for all $n\leq 0$, then $v(x)$ is invisible from the left for all wavenumbers that are integer multiples of $\pi/a$. The $a\to\infty$ limit of this result suggests the full-band left-invisibility of any potential whose Fourier transform $\tilde v(\fK)$ vanishes for $\fK\leq 0$. Surprisingly this result holds true even for the potentials that are not weak \cite{horsley}, i.e., they enjoy perfect left-invisibility.

The vanishing of $\tilde v(\fK)$ for $\fK\leq 0$ is equivalent to requiring the real and imaginary parts of $v(x)$ to be related by the Kramers-Kronig relations \cite{horsley}. The potentials of this type are generally long-range and their Schr\"odinger equation might not admit asymptotically plane-wave (Jost) solutions. This in turn makes their physical realization more difficult and leads to problems with the application of the standard scattering theory \cite{longhi-epl}.

The condition, $c_{n}=0$ for $n\leq 0$, for the unidirectional invisibility of weak locally periodic potentials of the form (\ref{e1}) follows as a simple byproduct of a dynamical formulation of scattering theory where the transfer matrix of the potential is given by the solution of a dynamical equation \cite{ap-2014}. We have recently developed a multi-dimensional extension of this formulation \cite{pra-2016} and employed it in the study of unidirectional invisibility in two and three dimensions \cite{prsa-2016}. Here we use it as a basic framework for exploring finite-band invisibility in two dimensions.

\begin{figure*}[t]
    \vspace{12pt}
    \begin{center}
    \includegraphics[scale=0.55]{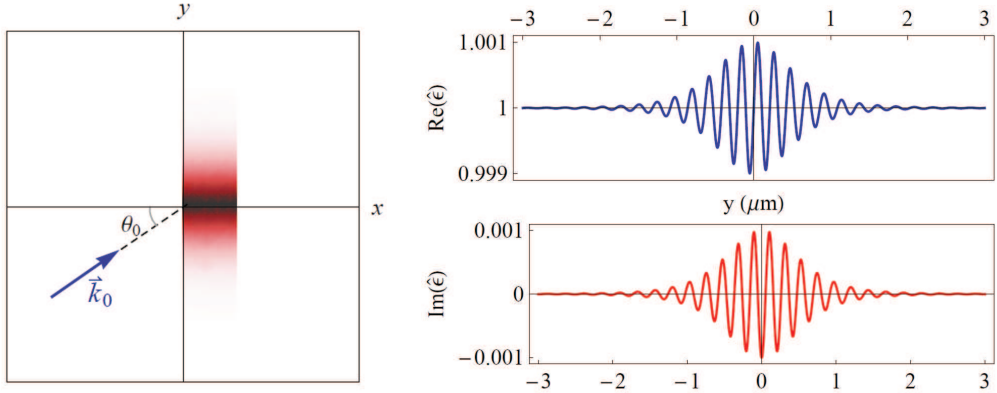}\\
    \caption{A schematic view of an oblique wave incident upon an inhomogeneous medium confined between the planes $x=0$ and $x=a$ (on the left), and plots of the real and imaginary parts of its relative permittivity $\hat\varepsilon$ that is given by (\ref{v=1}) and (\ref{ep=}) with $\alpha=2\pi/500~{\rm nm}$, $L=1~\mu{\rm m}$, $n=4$, and $x\in[0,a]$ (on the right).}
    \label{fig1}
    \end{center}
    \end{figure*}

Let $v(x,y)$ be a scattering potential in two dimensions, and suppose that the solutions of the Schr\"odinger equation
    \be
    -\nabla^2\psi(x,y)+v(x,y)\psi(x,y)=k^2\psi(x,y),
    \label{sch-eq}
    \ee
have the asymptotic form:
    \be
    \frac{1}{2\pi}\int_{-k}^k dp\, e^{ipy}\left[A_\pm(p)e^{i\omega(p)x}+
	B_\pm(p) e^{-i\omega(p)x}\right],
    \label{e3}
    \ee
for $x\to\pm\infty$, where $A_\pm(p)$ and $B_\pm(p)$ are functions vanishing for $|p|>k$, $\omega(p):=\sqrt{k^2-p^2}$, and the $x$-axis is the scattering axis. We can write the wavevector for a left-incident wave that makes an angle $\theta_0$ with the $x$-axis in the form $\vec k_0=
\omega(p_0)\hat e_x+p_0\hat e_y$, where $\hat e_x$ and $\hat e_y$ are respectively the unit vectors pointing along the $x$- and $y$-axes, and $p_0:=k\sin\theta_0$ (See Fig.~\ref{fig1}.) For such an incident wave,
$A_-(p)=2\pi\delta(p-p_0)$, $B_+(p)=0$, and the scattering solution of (\ref{sch-eq}) satisfies:
	\be
	\psi(\vec r)\to e^{i\vec k_0\cdot\vec r}+\sqrt{i/kr}\,e^{ikr} f(\theta)~~{\rm as}~~ r\to\infty,
	\nn
	\ee
where $\vec r:=x\hat e_x+y\hat e_x$, $(r,\theta)$ are the polar coordinates of $\vec r$, and $f(\theta)$ is the scattering amplitude.

The transfer matrix of the potential $v(x,y)$ is the $2\times 2$ matrix $\bM(p)$ fulfilling
    \be
	\bM(p)\left[\begin{array}{c} A_-(p)\\ B_-(p)\end{array}\right]=
	\left[\begin{array}{c} A_+(p)\\ B_+(p)\end{array}\right].
	\nn
	\ee
Its entries $M_{ij}(p)$ are linear operators acting on the functions $A_-(p)$ and $B_-(p)$. In Ref.~\cite{pra-2016} we show that $\bM(p)$ stores all the information about the scattering features of $v(x,y)$. In particular, if we set $T_-(p):=B_-(p)$ and $T_+(p):=A_+(p)-A_-(p)$, we can show that
    \begin{align}
	&T_-(p)=-2\pi M_{22}(p)^{-1}M_{21}(p)\delta(p-p_0),
    	\label{Tm-L}\\
	&T_+(p)=M_{12}(p)T_-(p)+2\pi[M_{11}(p)-1]\delta(p-p_0),
   	\label{Tp-L}\\
	&f(\theta)=-\frac{ik|\cos\theta|}{\sqrt{2\pi}}\times \left\{
	\begin{array}{cc}
	T_-(k\sin\theta) &{\rm for}~ \cos\theta<0\\
	T_+(k\sin\theta) &{\rm for}~  \cos\theta\geq 0
	\end{array}\right..
	\label{f=}
	\end{align}

A practically important property of the transfer matrix $\bM(p)$ is that it has the same composition property as its one-dimensional analog \cite{pra-2016}. This follows from the remarkable fact that
    \be
    \bM(p)=\bU(\infty,p),
    \label{M=}
    \ee
where $\bU(x,p)$ is the evolution operator for an effective non-Hermitian Hamiltonian operator $\bH(x,p)$ with $x$ playing the role of an evolution parameter. To make this statement more precise, we first introduce $v(x,i\partial_p)$ as the operator defined by
    \be
    v(x,i\partial_p)\phi(p):=\frac{1}{2\pi}\int_{-k}^k dq\,\tilde v(x,p-q) \phi(q),
	\label{v-dp}
	\ee
where $\phi(p)$ is a test function vanishing for $|p|>k$, and $\tilde v(x,\fK_y)$ is the Fourier transform of $v(x,y)$ with respect to $y$, i.e.,
    \be
    \tilde v(x,\fK_y):=\int_{-\infty}^\infty dy\,e^{-i\fK_y y}v(x,y).
    \label{FT}
    \ee
Equation (\ref{M=}) holds provided that we identify $\bU(x,p)$ with the solution of
    \be
    i\partial_x\bU(x,p)=\bH(x,p)\bU(x,p),~~~~\bU(-\infty,p)=\bI,
    \label{sch-eq-2}
    \ee
where
    \be
	\bH(x,p):=\frac{1}{2\omega(p)}\: e^{-i\omega(p)x\boldsymbol{\sigma}_3}
	v(x,i\partial_p)\,\boldsymbol{\cK}\,e^{i\omega(p)x\boldsymbol{\sigma}_3},
    	\label{H=}
    	\ee
$\bI$ is the $2\times 2$ identity matrix, $\boldsymbol{\sigma}_i$ are the Pauli matrices, and $\boldsymbol{\cK}:=\boldsymbol{\sigma}_3+i\boldsymbol{\sigma}_2$, \cite{pra-2016}.

It is important to realize that all the quantities we have introduced, in particular $\bM(p)$ and $\bH(x,p)$, depend on the wavenumber $k$. If $\bH(x,p)$ equals the zero operator $\bzero$ for a value of $k$, then (\ref{M=}) and (\ref{sch-eq-2}) imply $\bM(p)=\bI$ for this value of $k$. In light of (\ref{Tm-L}), (\ref{Tp-L}), and (\ref{f=}), this gives $f(\theta)=0$ for all $\theta_0$, i.e., the potential is invisible for any incident plane wave with this wavenumber. Because this argument does not rely on any approximation, this invisibility is perfect. Furthermore if this property holds for a range of values of $k$, then the potential will be perfectly invisible for any wave packet that is constructed by superposing the plane waves with wavenumber belonging to this range.

Now, suppose that there is some $\alpha>0$ such that $\tilde v(x,\fK_y)=0$ for all $\fK_y\leq 2\alpha$. Then in view of (\ref{v-dp}) and (\ref{H=}), $\bH(x,p)=\bzero$ for all $k\leq\alpha$, and the argument of the preceding paragraph proves the following result.
    \begin{itemize}
    \item[] {\em Theorem~1:} Let $\alpha>0$ and $v(x,y)$ be a scattering potential such that
    \be
    \tilde v(x,\fK_y)=0~~\mbox{for all}~~\fK_y\leq2\alpha.
    \label{condi}
    \ee
    Then $v(x,y)$ is perfectly invisible for any incident plane wave with wavenumber $k\leq\alpha$.
    \end{itemize}
According to this theorem, we can achieve perfect invisibility in the spectral band $[0,\alpha]$, if we can construct a potential satisfying (\ref{condi}). This is actually quite easy. With the help of (\ref{FT}), we can express every such potential in the form
    \be
    v(x,y)=e^{i2\alpha y}u(x,y),
    \label{v=u}
    \ee
where $u(x,y)$ satisfies $\tilde u(x,\fK_y)=0$ for all $\fK_y\leq 0$, i.e., for each fixed value of $x$, $u_x(y):=u(x,y)$ is one of the potentials considered in \cite{horsley,longhi-epl,longhi-ol-2015,longhi-ol-2016}. Clearly,
    \be
    u(x,y)=\frac{1}{2\pi}\int_0^\infty dq e^{iqy}\tilde u(x,q),
    \label{u=}
    \ee
where
for $x\to\pm\infty$, $|\tilde u(x,q)|\to 0$ sufficiently fast so that the solutions of (\ref{sch-eq}) have the asymptotic expression~(\ref{e3}). This condition is clearly satisfied for
    \be
    \tilde u(x,q)=\chi_a(x)\tilde f(x,q),~~~~q\geq 0,
    \label{u=1}
    \ee
where $\chi_a$ is the function defined in (\ref{e1}) and $\tilde f$ is an arbitrary function fulfilling $\int_0^\infty dq|\tilde f(x,q)|<\infty$. As an example, let $\tilde f(x,q)=\tilde\fz\, e^{-L q}q^n$, where
$\tilde\fz$ and $L$ are real parameters, $L>0$, and $n$ is a nonnegative integer. Then
(\ref{u=}) and (\ref{u=1}) give
    \be
    u(x,y)= \fz\, \chi_a(x)\left(\frac{y}{L}+i\right)^{-n-1} ,
    \label{v=1}
    \ee
where $\fz:=n!\,\tilde\fz/2\pi(-iL)^{n+1}$. Note that for $|y|\to\infty$, $|v(x,y)|\propto |L/y|^{n+1}$.

Next, we explore optical realizations of the perfect invisibility discussed in Theorem~1. Consider a nonmagnetic optical medium with translational symmetry along the $z$-axis, so that its properties are described by a relative permittivity $\hat\varepsilon$ that depends only on $x$ and $y$. A $z$-polarized TE wave propagating in this medium has an electric field of the form $\vec E(x,y,z)=E_0\,e^{-ik{\rm c}t}\psi(x,y)\hat e_z$, where $E_0$ is a constant, $c$ is the speed of light in vacuum, $\hat e_z$ is the unit vector pointing along the $z$-axis, and $\psi$ solves the Helmholtz equation $[\nabla^2+k^2\hat\varepsilon(x,y)]\psi=0$. The equivalence of this equation and the Schr\"odinger equation (\ref{sch-eq}) for the optical potential:
    \be
    v(x,y)=k^2[1-\hat\varepsilon(x,y)],
    \label{op-v}
    \ee
together with Theorem~1 prove the following result.
    \begin{itemize}
    \item[]{\em Theorem~2:} Let $u(x,y)$ be a function such that $\tilde u(x,\fK_y)=0$ for $\fK_y\leq 0$. Then a nonmagnetic optical medium described by the permittivity profile
    \be
    \hat\varepsilon(x,y)=1+e^{2i\alpha y} u(x,y),
    \label{ep=}
    \ee
is perfectly invisible for any incident TE wave with wavenumber $k\leq \alpha$.
    \end{itemize}
In particular, if (\ref{u=}) and (\ref{u=1}) hold, (\ref{ep=}) describes an optical slab of thickness $a$ that is invisible for these waves.

Next, consider choosing $\tilde f$ in (\ref{u=1}) in such a way that $u(x,y)$ decays rapidly for $y\to\pm\infty$. Then (\ref{ep=}) describes a slab of finite extension along both $x$- and $y$-axes. For example, the permittivity profile (\ref{ep=}) with $u(x,y)$ given by (\ref{v=1}) models a slab with a rectangular cross section,
    \be
    D=\{(x,y)| x\in[0,a], y\in [-b,b]\},
    \label{D}
    \ee
provided that $(L/b)^{n+1}\ll 1$. Figure~\ref{fig1} shows a plot of the real and imaginary parts of $\hat\varepsilon(x,y)$ for $u(x,y)$ given by (\ref{v=1}), $\alpha=2\pi/500~{\rm nm}$, $\fz=10^{-3}$, $L=1~\mu{\rm m}$, and $n=4$. These values yield $|\hat\varepsilon(x,y)-1|<7\times 10^{-6}$ for $|y|>2.5~\mu{\rm m}$. Therefore, we can use $\hat\varepsilon(x,y)$ to model a slab of cross section $D$ with $b\geq 2.5~\mu{\rm m}$, which is invisible for TE waves of wavelength $\lambda:=2\pi/k\geq 500~{\rm nm}$. In Supplementary Materials we show that the error one makes by modeling such a slab using this choice for $\hat\varepsilon(x,y)$ is indeed negligible.

In order to provide a graphical demonstration of the invisibility of the above system for TE waves, we compute its scattering amplitude using the first Born approximation. This is a reliable approximation scheme, because the corresponding optical potential is sufficiently weak.

Performing the first Born approximation corresponds to solving the dynamical equation (\ref{sch-eq-2}) for the transfer matrix using first-order perturbation theory, i.e.,
$\bM(p)\approx \bI-i\int_{-\infty}^\infty dx\bH(x,p)$, \cite{pra-2014a,prsa-2016}.
Substituting (\ref{H=}) in this equation, we obtain explicit formulas for the action of $M_{ij}(p)$ on test functions $\phi(p)$
. These together with (\ref{Tm-L}) -- (\ref{f=}) imply
    \be
    f(\theta)\approx\frac{-1}{2\sqrt{2\pi}}\,\tilde{\tilde v}\big(k(\cos\theta-\cos\theta_0),
    k(\sin\theta-\sin\theta_0)\big),
    \label{f=2}
    \ee
where $\tilde{\tilde v}(\fK_x,\fK_y):=\int_{-\infty}^\infty \!\!dx\int_{-\infty}^\infty\!\! dy\,e^{-i(\fK_xx+\fK_yy)}v(x,y)$ is the two-dimensional Fourier transform of $v(x,y)$.

For a scattering potential $v(x,y)$ satisfying $\tilde v(x,\fK_y)=0$ for $\fK_y\leq 2\alpha$, we have $\tilde{\tilde v}(\fK_x,\fK_y)=0$ for $\fK_y\leq 2\alpha$. In view of this observation and the fact that $|\sin\theta-\sin\theta_0|\leq 2$, the right-hand side of (\ref{f=2}) vanishes. This provides a first-order perturbative verification of Theorem~1, which actually holds to all orders of perturbation theory.

\begin{figure}[ht]
    \vspace{12pt}
    \begin{center}
    \includegraphics[scale=.4]{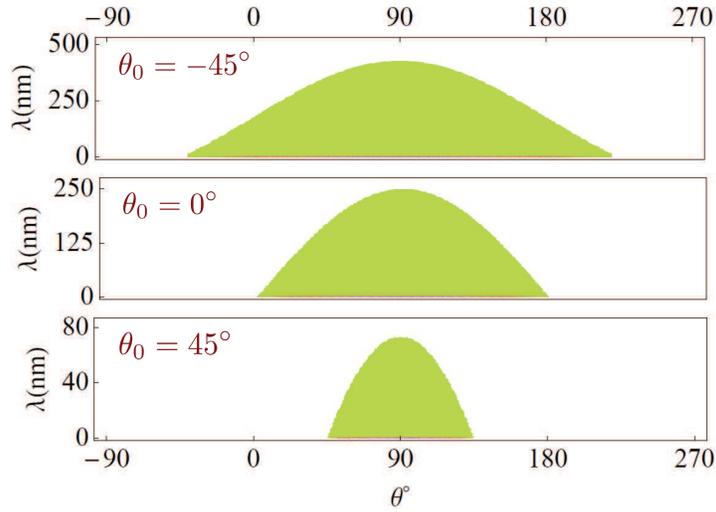}
    \caption{{Visibility domains of the permittivity profile ${\hat\varepsilon=1+
    \fz\, \chi_a(x)e^{2i\alpha y}\left({y}/{L}+i\right)^{-5}}$ for TE waves:} The colored regions correspond to values of $\lambda$ and $\theta$ for which $f(\theta)\neq 0$. The top, middle, and bottom graphs correspond to TE waves with incidence angle $\theta_0=-45^\circ, 0^\circ$, and $45^\circ$, respectively. The parameters specifying $\hat\varepsilon$ are taken as: $\fz=10^{-3}$, $a=100~\mu{\rm m}$, $L=1~\mu{\rm m}$, and $\alpha=2\pi/500~{\rm nm}$.
    For all values of $\theta_0$ the system is invisible for $\lambda\geq 500~{\rm nm}$. As one increases $\theta_0$, the system becomes invisible above a critical wavelength that is smaller than  $500~{\rm nm}$.}
    \label{fig2}
    \end{center}
    \end{figure}

We can use (\ref{f=2}) to determine the wavelengths $\lambda$ at which a weak optical potential is invisible for TE waves. Figure~\ref{fig2} shows regions in the $\theta$-$\lambda$ plane where $f(\theta)\neq 0$ for some TE waves that propagate in a medium with permittivity profile given by (\ref{v=1}), (\ref{ep=}), $\fz=10^{-3}$, $a=100~\mu{\rm m}$, $L=1~\mu{\rm m}$, $\alpha=2\pi/500~{\rm nm}$, and $n=4$. This profile, which can be realized using a slab of thickness $a=100~\mu{\rm m}$ and width $2b\geq 5~\mu{\rm m}$ placed in vacuum, is invisible for the TE waves with an arbitrary incidence angle $\theta_0$ and wavelength $\lambda\geq 500~{\rm nm}$.

Next, we study the propagation of a TM wave in a nonmagnetic isotropic medium described by a relative permittivity profile $\hat\varepsilon(x,y)$. The magnetic field for this wave has the form $\vec H(x,y,z)=H_0\,e^{-ik{\rm c}t}\phi(x,y)\hat e_z$, where $H_0$ is a constant and $\phi$ is a function. Imposing Maxwell's equations, we find that $\phi$ satisfies
    \be
    \hat\varepsilon^{-1}\nabla^2\phi+
    \vec\nabla(\hat\varepsilon^{-1})\cdot\vec\nabla\phi+k^2\phi=0.
    \label{TM-eqn}
    \ee
This becomes equivalent to the Schr\"odinger equation (\ref{sch-eq}) provided that we set $\psi:=\phi/\sqrt{\hat\varepsilon}$,
    \bea
    v:=-k^2\eta +\frac{3|\vec\nabla\eta |^2}{4(1+\eta)^2}-
    \frac{\nabla^2\eta }{2(1+\eta)},
    \label{TM-v}
    \eea
and $\eta:=\hat\varepsilon-1$. For a permittivity profile of the form (\ref{ep=}),
$\eta(x,y)=e^{2i\alpha y} u(x,y)$. Therefore, $\tilde\eta(x,\fK_y)=0$ for $\fK_y\leq 2\alpha$ provided that $\tilde u(x,\fK_y)=0$ for $\fK_y\leq 0$.

In Supplementary Materials we show that if $\hat\varepsilon$ is bounded and its real part exceeds a positive value, i.e., there are positive real numbers $m$ and $M$ such that $m\leq \RE(\hat\varepsilon)\leq|\hat\varepsilon|\leq M$, then the vanishing of $\tilde\eta(x,\fK_y)$ for $\fK_y\leq 2\alpha$ implies that the same holds for the potential (\ref{TM-v}). Therefore it satisfies the invisibility condition (\ref{condi}), and we are led to the following result.
    \begin{itemize}
    \item[]{\em Theorem~3:} A nonmagnetic optical medium described by a smooth relative permittivity profile of the form $\hat\varepsilon(x,y)=1+e^{2i\alpha y} u(x,y)$ is perfectly invisible for incident TM waves of wavenumber $k\leq\alpha$ provided that $\tilde u(x,\fK_y)=0$ for $\fK_y\leq 0$ and there are positive numbers $m$ and $M$ such that $m\leq\RE[\hat\varepsilon(x,y)]\leq|\hat\varepsilon(x,y)|\leq M$.
    \end{itemize}
If the hypothesis of this theorem holds except that we allow $u$ to have discontinuities along boundaries of certain connected regions $D_\alpha$, then we need to solve (\ref{TM-eqn}) in $D_\alpha$ and patch the solutions for adjacent $D_\alpha$ by imposing the standard electromagnetic interface conditions along their common boundaries. Because the interface conditions involve $\hat\varepsilon$, the presence of the discontinuities can make the system visible even for $k\leq \alpha$. This is true unless the resulting jumps in the value of $|\hat\varepsilon|$ are negligibly small. For example consider the optical slab we examined in our discussion of the TE waves, and suppose that $u(x,y)$ is given by the right-hand side of (\ref{v=1}) multiplied by $g(x)=e^{-(2x-a)^2/\sigma^2}$. Then, for $\sigma\ll a$, we can safely ignore the contribution of the discontinuity of $\hat\varepsilon$ along the boundaries of the slab and conclude that it is practically invisible for both TE and TM waves with $k\leq \alpha$.

In summary, we have introduced a simple criterion for perfect finite-band invisibility in two dimensions and explored some of its optical realizations. In contrast to the criterion of Ref.~\cite{horsley} for the full-band unidirectional invisibility in one dimension, ours does not restrict the asymptotic decay rate of the potential along the scattering axis. This is a key feature of our route to broadband invisibility that allows for its realization using optical slabs. Furthermore, the width of the spectral band in which the invisibility is effective is a free parameter in our construction. Finally, we would like to mention that our results easily extend to three dimensions. This is because the dynamical formulation of scattering in three dimensions \cite{pra-2016} involves an effective Hamiltonian operator with the same structure as its two-dimensional analog. We expect the resulting broadband invisibility in three dimensions to find interesting applications in acoustics.\\[3pt]

\noindent \textbf{Acknowledgments:} This project was supported by the Turkish Academy of Sciences (T\"UBA).

{

}

\np

\begin{center}
{\bf \large Supplementary Material}
\end{center}

\noindent\textbf{A. Using (\ref{v=1}) and (\ref{ep=}) to describe a slab with a rectangular cross section}\\[3pt]

Let $v_0(x,y)$ be the optical potential (\ref{op-v})  for the permittivity profile given by  (\ref{v=1}) and (\ref{ep=}), i.e.,
        \bea
	    v_0(x,y)&:=&-k^2\fz\chi_a(x)e^{2i\alpha y}
	    \left(\frac{y}{L}+i\right)^{-n-1}.
        \nn
        \eea
Truncating this potential for $|y|\geq b$ yields the potential,
        \bea
	    v(x,y)&=&\left\{\begin{array}{cc}
        v_0(x,y)&{\rm for}~|y|\leq b,\\
        0 &{\rm otherwise},\end{array}\right.\nn
        \eea
which describes a slab with a rectangular cross section (\ref{D}).
	
It is easy to show that
    \bea
    \tilde{\tilde v}(\fK_x,\fK_y)&=& X(\fK_x) Y_n(\fK_y-2\alpha,\ell),
    \nn
    \eea
where
    \begin{align}
    &X(q):=ik^2L\fz q^{-1}(1-e^{-iaq}),
    &&Y_n(q,\ell):=\int_{-1/\ell}^{1/\ell} dt\; \frac{e^{-iqLt}}{\left(t+i\right)^{n+1}},
    &&\ell:=\frac{L}{b}.\nn
    \end{align}

For the particular values of the parameters of the system that we consider, both $v_0(x,y)$ and $v(x,y)$ are weak enough for the first Born approximation to be reliable. According to Eq.~(\ref{f=2}), i.e.,
    \be
    f(\theta)\approx\frac{-1}{2\sqrt{2\pi}}\,\tilde{\tilde v}\big(k(\cos\theta-\cos\theta_0),
    k(\sin\theta-\sin\theta_0)\big),
    \nn
    \ee
and the fact that $\tilde{\tilde v}_0(\fK_x,\fK_y)=X(\fK_x) Y_n(\fK_y-2\alpha,0)$, the error we make by using $v_0(x,y)$ instead of $v(x,y)$ for computing the scattering amplitude is proportional to
    \bea
    \left|Y_n(q,\ell)-Y_n(q,0)\right|&=&
        \left|\int_{-\infty}^{-1/\ell} dt\;\frac{e^{-iqLt}}{\left(t+i\right)^{n+1}}+
              \int_{1/\ell}^\infty dt\;\frac{e^{-iqLt}}{\left(t+i\right)^{n+1}}\right|\nn\\
        &=&\left|\int_{1/\ell}^\infty dt\;
                \left[\frac{e^{-iqLt}}{\left(t+i\right)^{n+1}}+
                \frac{e^{iqLt}}{\left(-t+i\right)^{n+1}}\right]\right|\nn\\
                     &\leq&\int_{1/\ell}^\infty dt\;
                    \left(\left|\frac{e^{-iqLt}}{\left(t+i\right)^{n+1}}\right|+
                    \left|\frac{e^{iqLt}}{\left(-t+i\right)^{n+1}}\right|\right)=
                    2\int_{1/\ell}^\infty dt\;(t^2+1)^{-\frac{n+1}{2}}\nn \\
                    &\leq& 2\int_{1/\ell}^\infty dt\; t^{-(n+1)}=\frac{2\ell^n}{n}.\nn
    	     \eea
For the system we consider, $L=1\,\mu{\rm m}$, $n=4$, and $b\geq 2.5\,\mu{\rm m}$. These in turn imply $\ell=L/b\leq 0.4$ and $2 \ell^n/n<0.013$. In particular, for $b=5\,\mu{\rm m}$, we have $2 \ell^n/n=8
\times 10^{-4}$. Therefore,  for this value of $b$, we can safely use $v_0(x,y)$ to describe the scattering effects of the slab. 
\pagebreak

\noindent \textbf{B. Certain properties of functions $\boldsymbol{u(x,y)}$ satisfying $\boldsymbol{\tilde u(x,\fK_y)=0}$ for $\boldsymbol{\fK_y\leq 0}$ and the invisibility for TM waves}\\[3pt]

Suppose that $u_j(x,y)$ with $j=1,2$ be a pair of functions with this property, and $w(x,y):=u_1(x,y)u_2(x,y)$. Then, we have
        \begin{align}
        &\widetilde{\partial_x u_j}(x,\fK_y)=\partial_x\tilde u_j(x,\fK_y)=0,~~~~~\mbox{for $\fK_y\leq 0$},
        \label{i}\\
        &\widetilde{\partial_y u_j}(x,\fK_y)=-i\fK_y\tilde{u}_j(x,\fK_y)=0, ~~~~~\mbox{for $\fK_y\leq 0$}.
        \label{ii}
        \end{align}
    Furthermore, by virtue of the convolution formula,
        \be
        \tilde w(x,\fK_y)=\int_{-\infty}^\infty dq\:\tilde u_1(x,\fK_y-q)\tilde u_2(x,q)=\int_0^\infty dq\:\tilde u_1(x,\fK_y-q)\tilde u_2(x,q).
        \label{iiia}
        \ee
    For $\fK_y\leq 0$ and $q\geq 0$, $\fK_y-q\leq 0$. This shows that $\tilde u_1(x,\fK_y-q)=0$, and, in view of (\ref{iiia}),
        \be
        \tilde w(x,\fK_y)=0,~~~~~\mbox{for $\fK_y\leq 0$}.
        \label{iii}
        \ee

    Next, let $\alpha>0$, $u(x,y)$ be a function such that $\tilde u(x,\fK_y)=0$ for $\fK_y\leq 0$, $\eta(x,y):=e^{2i\alpha y}u(x,y)$, and $\xi(x,y)$ be a function that can be written as a sum of terms of the form $\eta^\ell(\partial^m_i\eta)(\partial^n_j\eta)$ where $\ell,m$, and $n$ are nonnegative integers, $i,j\in \{x,y\}$, and $\partial^0_i\eta:=\eta$. Then (\ref{i}), (\ref{ii}), and the argument leading to (\ref{iii}) imply that $\xi(x,y)=e^{2(\ell+2)i\alpha y}u_3(x,y)$ for some function $u_3(x,y)$ such that $\tilde u_3(x,\fK_y)=0$ for $\fK_y\leq 0$. This gives $\tilde\xi(x,\fK_y)=0$ for $\fK_y\leq 2(\ell+2)\alpha$. Because $\ell+2\geq 1$, we have
        \be
        \tilde\xi(x,\fK_y)=0~~~~~\mbox{for $\fK_y\leq2\alpha$.}
        \label{txi=}
        \ee
    We can use this result to show that if there are positive real numbers $m$ and $M$ such that
        \be
        m\leq\RE[\hat\varepsilon(x,y)]\leq|\hat\varepsilon(x,y)|\leq M,
        \label{e-bnd}
        \ee
     then the optical potential (\ref{TM-v}) for the TM waves, namely
        \be
        v:=-k^2\eta +\frac{3|\vec\nabla\eta |^2}{4(1+\eta)^2}-
        \frac{\nabla^2\eta }{2(1+\eta)},
        \label{v-TM}
        \ee
    satisfies
        \be
        \tilde v(x,\fK_y)=0~~\mbox{for all}~~\fK_y\leq2\alpha.
        \label{condi}
        \ee
    To do this, we first introduce
        \begin{align}
        &\beta(x,y):=\frac{\eta(x,y)-\mu}{1+\mu}, &&\mu:=\frac{M^2+1}{2m},
        \nn
        \end{align}
    and use them to establish the identity:
        \be
        \frac{1}{(1+\eta)^s}=\frac{1}{(1+\mu)^s}\frac{1}{(1+\beta)^s},
        \label{exp1}
        \ee
    where $s=1,2$. The right-hand side of (\ref{exp1}) admits a convergent power series in $\beta$ provided that $|\beta|<1$. We next show that this condition is indeed fulfilled.

    Because $\eta=\varepsilon-1$, we can use (\ref{e-bnd}) to show that
        \bea
        |\eta|^2&\leq&|\varepsilon|^2+1\leq M^2+1\nn\\
        &\leq&\frac{M^2\RE(\varepsilon)}{m}+1=(2\mu -\frac{1}{m})\RE(\varepsilon)+1 \nn\\
        &<&2\mu \RE(\varepsilon)+1=2\mu \RE(\eta)+ 2\mu+1.\nn
        \eea
    This in turn implies
        \bea
        |\beta|= \frac{|\eta-\mu|}{1+\mu} =
        \frac{\sqrt{[\RE(\eta)-\mu]^2+\IM(\eta)^2}}{1+\mu}=
        \frac{\sqrt{|\eta|^2 -2\mu\RE(\eta)+\mu^2}}{1+\mu}<1.
        \label{bnd-2}
        \eea
    This relation ensures that the binomial series expansion of the second factor on the right-hand side of (\ref{exp1}) converges to the value of this factor, i.e.,
        \be
        \frac{1}{(1+\eta)^s}=\frac{1}{(1+\mu)^s}\sum_{n=0}^\infty n^{s-1}(-\beta)^{n-s+1},~~~~~~~~~~s=1,2.
        \label{exp2}
        \ee

    To prove that (\ref{v-TM}) satisfies (\ref{condi}) we only need to show that the latter holds for the potential
        \be
        v=\frac{3|\vec\nabla\eta |^2}{4(1+\eta)^2}-\frac{\nabla^2\eta }{2(1+\eta)}.
        \label{v2}
        \ee
    We can use (\ref{exp2}) to expand the right-hand side of (\ref{v2}) in a convergent power series in $\beta$. Each term of this series is a linear combination of terms of the form $\eta^\ell(\partial^m_i\eta)(\partial^n_j\eta)$ with $m+n=2$ and $\ell\geq 0$, i.e., it is an example of the function $\xi$ that satisfies (\ref{txi=}). This proves (\ref{condi}) for the potential (\ref{v2}).

\end{document}